# Delay-Tolerant Networking for Tsunami Evacuation on the Small Island of Hachijojima: A Study of Epidemic and Prophet Routing

Keiya Kawano
*School of Computer Science*
*The University of Nottingham*
psxkk12@nottingham.ac.uk

Milena Radenkovic
*School of Computer Science*
*The University of Nottingham*
milena.radenkovic@nottingham.ac.uk

*Abstract*— Tsunami disasters pose a serious and recurring threat to coastal and island communities. When a large earthquake occurs, people are forced to make evacuation decisions under extreme time pressure, often at the same time as the communication infrastructure is damaged or completely lost. In such circumstances, the familiar channels for sharing information - cellular networks, the internet, and even landlines - can no longer be relied upon. What typically remains are the mobile devices that evacuees carry with them. These devices can form Delay Tolerant Networks (DTNs), in which messages are forwarded opportunistically whenever people come into contact. To explore this, we evaluate multi-criteria performance characteristics of two DTN routing schemes in a pre-tsunami evacuation scenario for the island of Hachijojima, Japan use case.

## I. Introduction

### A. Introduction

Tsunami disasters pose a serious and recurring threat to coastal and island communities [1]. When a large earthquake occurs, people are forced to make evacuation decisions under extreme time pressure, often at the same time as the communication infrastructure is damaged or completely lost. In such circumstances, the familiar channels for sharing information - cellular networks, the internet, and even landlines - can no longer be relied upon. What typically remains are the mobile devices that evacuees carry with them. These devices can form Delay Tolerant Networks (DTNs), in which messages are forwarded opportunistically whenever people come into contact. In this work, we are less concerned with modelling the fine details of human evacuation behaviour and more interested in a complementary question: when infrastructure is unreliable and human movement is largely uncoordinated, how well can DTN routing protocols deliver critical information to safe locations? To explore this, we evaluate two widely studied DTN routing schemes - Epidemic routing and PROPHET- in a tsunami pre-evacuation scenario. Our study focuses on Hachijojima, a small island south of Tokyo with a geographically constrained road network and limited alternative routes. Using THE ONE simulator, we model three types of mobile nodes- pedestrians, cars, and emergency vehicles - that move independently along the road network according to a shortest-path map-based mobility model. Importantly, these nodes are not explicitly guided to shelters in the mobility model; they wander the road network without "knowing" where shelters are. Shelters, instead, are modelled as stationary, high-capacity DTN nodes placed at safe, elevated points on the island. All application-layer messages are addressed to these shelter nodes, which serve as logical sinks and aggregation points for information. By stress-testing Epidemic and PROPHET routing under this deliberately simplified and pessimistic mobility assumption, we aim to understand which protocol offers better delivery delay and delivery ratio to shelter nodes within a limited time window before the tsunami arrives. The broader goal of this study is to provide insight into how resilient communication mechanisms can be designed for island evacuations, where infrastructure damage is likely, movement is messy rather than perfectly coordinated, and yet the timely arrival of information at safe hubs can still make a difference.

### B. Background

Japan is one of the most seismically active regions in the world, and large earthquakes capable of generating tsunamis are not rare events. Historical data, such as that summarised in Fig. 1, shows repeated occurrences of strong earthquakes between 1926 and 2008.

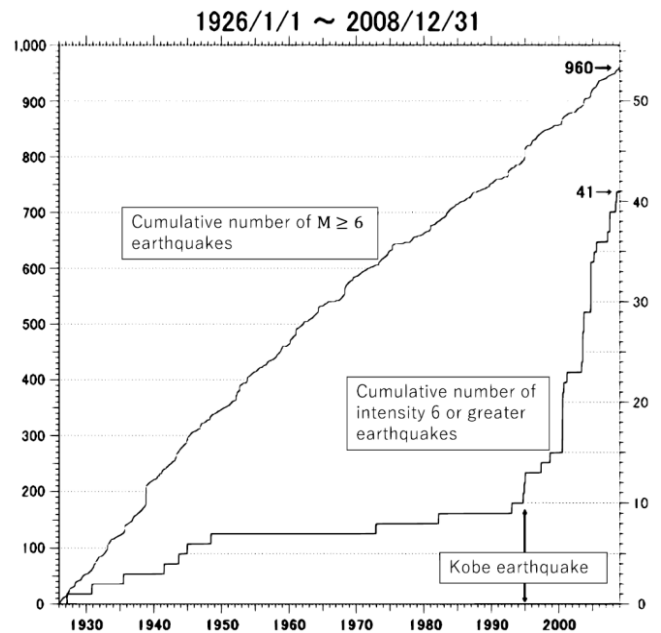

Fig. 1. The cumulative number of earthquakes of magnitude 6 or greater and seismic intensity 6 or greater occurring around Japan during the 83-year period from 1926 to 2008 [3]

When a tsunami is triggered, the height of the incoming waves is a key factor: once tsunami heights exceed roughly two metres, the rate at which houses and infrastructure are destroyed increases sharply, and the consequences quickly become catastrophic [2]. This implies a very simple but unforgiving requirement: both physical evacuation and the dissemination of critical information should be largely completed before such waves make landfall. At the same time, major earthquakes often damage the very infrastructure that communities rely on to coordinate their response. Power outages, broken backhaul links, and damaged base stations can all contribute to the partial or complete collapse of conventional communication networks. Under these conditions, the ability to form local, infrastructure-independent communication paths becomes crucial. DTNs, formed opportunistically by mobile devices carried by evacuees and emergency responders, offer one such mechanism: messages can be stored, carried, and forwarded as people move, even if end-to-end paths never exist at any single moment. Small islands are particularly vulnerable in this context. Their road networks are limited, evacuation routes are easily congested, and the population is often ageing [4]. Hachijojima, located at the southern edge of the Tokyo metropolitan area, is a representative example. The island has around 7,000 residents, and a significant proportion are elderly, which makes rapid, self-directed evacuation more difficult in practice. Its geographic isolation also means that help and backup infrastructure cannot simply be brought in immediately from neighbouring regions. In this report, we therefore take Hachijojima as a case study and construct a tsunami scenario in which evacuees move independently along the island's road network. Rather than enforcing that all evacuees physically converge on shelters, we treat shelters as stationary DTN nodes that act as safe information hubs. Our focus is on whether tsunami-related messages - alerts, local damage observations, and status updates - can be delivered to these shelter nodes within the expected tsunami arrival time after an earthquake near Hachijojima. Framing the problem in this way allows us to evaluate DTN routing protocols as a support mechanism for tsunami response, while acknowledging that the underlying human evacuation behaviour is represented in a simplified, map-based form rather than a fully realistic behavioural model.

## II. Delay Tolerant Networks (DTN)

### A. Overview

Delay Tolerant Networks (DTNs) are designed to operate in environments where end-to-end connectivity cannot be guaranteed, delays are substantial, and network partitions frequently occur. Unlike traditional IP networks, DTNs rely on a store–carry–forward paradigm, where nodes buffer messages, physically carry them while moving, and forward them opportunistically when encountering other nodes [5]. This architecture makes DTNs particularly suitable for emergency and disaster scenarios, such as tsunami evacuations, where infrastructure-based communication systems may fail or become overloaded [6] [7]. Routing in DTNs is inherently challenging because forwarding decisions must be made without reliable knowledge of the current network state, future mobility patterns, or contact opportunities. Numerous DTN routing strategies have been proposed, broadly categorised into flooding-based, probabilistic, history-aware, and social-aware approaches [8].

In this study, we focus on two of the most established and influential protocols: Epidemic Routing and PROPHET Routing, representing the extremes of the flooding- and prediction-based categories, respectively. Their contrasting mechanisms enable meaningful performance comparisons under the constrained and highly directional mobility patterns of evacuation scenarios.

### B. Epidemic Routing

Epidemic routing, first proposed by Vahdat and Becker (2000), is one of the earliest and most fundamental routing protocols for delay-tolerant networks. It follows a pure flooding strategy in which messages are aggressively replicated across the network. Each node maintains a summary vector that records the identifiers of all messages currently stored. When two nodes come into contact, they exchange their summary vectors and transmit any messages that the other node does not already possess. This process repeats at every contact until all reachable nodes have obtained the message or the message expires due to its time-to-live (TTL). The mechanism can be summarised as follows. When node A encounters node B, they first exchange summary vectors describing their buffered messages. For each message M that is present at A but not at B, node A forwards a copy of M to node B. Subsequent contacts propagate additional copies further through the network, ideally allowing M to reach all nodes, including its intended destination [9]. Epidemic routing offers several advantages. Because messages are widely replicated, the probability that a message eventually reaches its destination is very high. In relatively dense networks, where nodes meet frequently, this replication also results in low delivery latency, as messages quickly spread through multiple paths. Moreover, the protocol does not rely on any knowledge or prediction of node mobility and therefore remains effective even when movement patterns are random or unknown. These benefits come at a substantial cost. Frequent exchanges of summary vectors and message copies cause very high communication overhead and can congest the wireless channel. Nodes rapidly accumulate large numbers of messages, which increases buffer usage and leads to buffer overflows. The continuous transmission and reception of message copies also consume significant energy, making the protocol poorly suited to battery-powered mobile devices such as smartphones carried by evacuees. In the context of tsunami evacuations, where contact opportunities may be sparse and unpredictable, but the priority is to disseminate life-critical information as reliably as possible, Epidemic routing can be viewed as an upper bound on delivery performance. However, its heavy resource consumption means that it is unlikely to be feasible for real-world deployment during a disaster, and it primarily serves as a reference point for evaluating more efficient protocols.

### C. Prophet Routing

The PROPHET (Probabilistic Routing Protocol using History of Encounters and Transitivity) protocol, proposed by Lindgren et al. (2003), aims to reduce the overhead of flooding-based schemes by exploiting regularities in node encounters. Instead of forwarding messages indiscriminately, PROPHET uses past contact patterns to estimate how likely a node is to deliver a message to a given destination. Routing decisions are then based on these delivery predictability values [10]. Each node maintains, for every possible destination b, a delivery predictability value $P(a, b)$ that reflects how likely node a is to deliver a message to b. Whenever two nodes meet, their predictability values are updated. Direct encounters increase $P(a, b)$ for the encountered node, under the assumption that nodes that have met once are more likely to meet again. PROPHET also applies a transitivity rule: if node A often meets B, and B often meets C, then A is assumed to have a non-negligible chance of eventually delivering messages to C via B, and P(A,C) is increased accordingly [10]. To ensure the protocol adapts to changing mobility patterns, all predictability values are gradually reduced over time through an ageing mechanism. Forwarding decisions are made opportunistically at each contact. When two nodes meet, they exchange their predictability tables. For each message destined for node d, a node forwards the message only if the other node has a higher delivery predictability for d. In this way, messages tend to move towards nodes that are, statistically, better positioned to deliver them to the destination. Compared with Epidemic routing, PROPHET offers several advantages. Because it forwards messages selectively, it generates substantially lower overhead and exerts less pressure on node buffers. This makes the protocol more scalable and more suitable for deployment on resource-constrained devices. However, these benefits are accompanied by some limitations. In highly dynamic or directional scenarios, where nodes do not repeatedly encounter the same peers, the historical encounter information on which PROPHET relies becomes less informative. As a result, delivery probability may degrade, and messages can experience higher latency because they are forwarded less aggressively. In tsunami evacuation scenarios, evacuees typically move simultaneously from coastal areas towards a limited number of inland shelters. This produces strongly unidirectional mobility with few repeated encounters. Such conditions violate PROPHET's core assumption that past encounters are a good predictor of future ones. Consequently, while PROPHET can significantly reduce overhead, it is at risk of failing to deliver a sufficient fraction of critical messages when rapid and widespread dissemination is required.

*D. Comparative Relevance of Epidemic and PROPHET in Disaster Scenarios*

Epidemic and PROPHET routing embody two contrasting design philosophies for delay-tolerant networking. Epidemic routing represents the flooding-based extreme: it forwards copies to almost every encountered node, prioritising delivery probability and low latency at the expense of efficiency. PROPHET, in contrast, is a prediction-based protocol that forwards messages only when the encountered node appears more promising according to its delivery predictability estimates, thereby emphasising efficiency and scalability. In qualitative terms, the main trade-offs can be summarised as follows. Epidemic routing achieves very high delivery probability and short delays in many settings, but suffers from extreme communication overhead, rapid buffer exhaustion, and high energy consumption. PROPHET substantially reduces overhead and buffer usage by forwarding selectively, which is advantageous for mobile devices and congested wireless channels. However, its performance is sensitive to the presence of stable and repeatable contact patterns; when mobility is highly directional or non-repetitive, its delivery probability can be significantly lower than that of Epidemic routing. Disaster scenarios such as tsunamis create precisely the kind of challenging conditions under which these trade-offs become critical. Networks are often highly partitioned and entirely infrastructure-less; mobility is driven by evacuation routes that funnel people along specific paths; and contact opportunities are sparse and short-lived. Under such stress, Epidemic routing provides a near-ideal reference in terms of reliability and latency but is impractical for real deployment because it consumes excessive bandwidth, storage, and energy.

PROPHET, on the other hand, offers a more realistic resource footprint but may fail to deliver enough messages when its mobility assumptions are violated. These contrasting behaviours make Epidemic and PROPHET a particularly informative pair of protocols to compare in the context of an island tsunami evacuation model. Their performance in our simulations reveals how different routing strategies balance reliability against efficiency under severe constraints, and thus provides insight into the design of practical DTN-based communication support for real-world evacuations.

III. TSUNAMI EVACUATION PROBLEM DESIGN AND SETUP

This section describes how the tsunami evacuation scenario for Hachijojima was modelled, how the simulations were configured in the ONE (Opportunistic Network Environment) simulator, and how the performance of the two DTN routing protocols - Epidemic and PROPHET – was evaluated.

*A. Tsunami evacuation scenario*

We model a pre-tsunami evacuation phase on Hachijojima, a small Japanese island with a compact, constrained road network and only limited areas of safe, elevated ground. The scenario captures the critical time window between the issuance of an official tsunami warning and the expected arrival of the first wave. Within this window, we assume that conventional communication infrastructure (cellular networks, Wi-Fi backhaul) is either severely degraded or completely unavailable, so that information dissemination must rely on opportunistic contacts between mobile devices carried by evacuees and responders. In the simulator, evacuees are represented as mobile nodes - pedestrians, cars, and emergency vehicles - that move independently along the island's road network according to a shortest-path map-based mobility model. Crucially, these nodes are not explicitly directed towards shelters by the mobility model; they simply traverse the road graph without prior knowledge of where shelters are located. In a real earthquake, considerable chaos is expected, so our scenario deliberately reflects a pessimistic, worst-case situation. Furthermore, as gathering information at evacuation centres facilitates rescue operations originating from there, the plan is to centralise information at these centres. Shelters themselves are modelled as stationary DTN nodes placed at safe, elevated locations and equipped with large buffers. All application-layer messages in the scenario are addressed to these shelter nodes, which therefore act as logical sinks and information aggregation points rather than as physical crowd-attractors.

Whenever two nodes come within wireless communication range, they may exchange messages according to the DTN routing protocol under test. The primary objective of our simulations is to evaluate how effectively Epidemic and PROPHET routing can deliver time-critical messages (e.g., evacuation instructions, local situation updates, or safety confirmations) from mobile evacuees to shelter nodes within the limited pre-tsunami time window. In other words, we focus on the informational aspect of evacuation - how quickly and reliably crucial data reaches safe hubs under uncoordinated, infrastructure-poor conditions - rather than on reproducing detailed human crowd dynamics. The focus of this work is on comparing protocol families rather than performing fine-grained parameter tuning.

*B. Simulation conditions*

All experiments were conducted using the ONE simulator, which is widely used for DTN and opportunistic networking research [11]. The physical environment is based on real map data for Hachijojima. OpenStreetMap (OSM) data [12] for the island was converted into a WKT (Well-Known Text) road network file (map.wkt), which is then loaded into ONE as the map-based movement layer:

```
MapBasedMovement.nrofMapFiles = 1
```

```
MapBasedMovement.mapFile1 = data/map.wkt
```

The simulation time is set to 1 800 seconds, corresponding to a 30-minute evacuation window:

```
Scenario.endTime = 1800
```

This timeframe is chosen to reflect recommendations that evacuation and the dissemination of critical information should be largely completed within roughly 30-60 minutes after a tsunami generating earthquake [13]. Unlike the previous version of the scenario, no additional warm-up period is used; nodes begin moving and exchanging messages from the start of the 30-minute window, which we interpret as the period between the official tsunami warning and the expected arrival of the first wave. The global update interval is set to 0.1 seconds,

```
Scenario.updateInterval = 0.1
```

which provides a reasonably fine temporal granularity without incurring excessive simulation overhead. A base configuration uses MovementModel.rngSeed = 1, and in our experiments, we vary this seed between 1 and 5 as described later.

```
MovementModel.rngSeed = 1(to 5)
```

Fig. 2 shows the road network of Hachijojima imported into the ONE simulator.

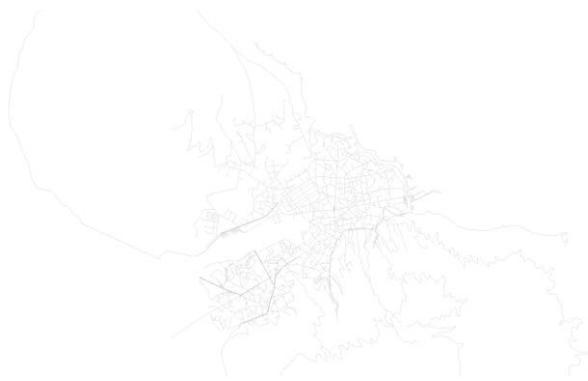

Fig. 2. The wkt file data of Hachijojima for simulation without node

*1) Choice of simulator.*

Several network simulators could in principle be used to model tsunami-evacuation DTNs, including general-purpose packet-level tools such as NS-3 and OMNeT++, as well as trace-driven frameworks that replay empirical contact logs. NS-3 and OMNeT++ offer detailed models of radio propagation, interference, and protocol stacks, but this level of detail comes at a substantial computational cost when simulating thousands of mobile nodes over many random seeds.

They also require significant additional effort to build message-oriented DTN routing logic and reporting scripts on top of their low-level primitives. In contrast, the ONE is explicitly designed for opportunistic and delay-tolerant networking: it provides built-in support for message-based DTN routing, map-based and trace-based mobility models, and ready-made reports for key performance metrics such as delivery probability, delay, and overhead. This makes it well-suited for rapid exploration of multiple routing protocols and scenario parameters [14]. The trade-off is that ONE's radio and MAC-layer models are deliberately simplified: interference, congestion, and physical-layer effects are abstracted into a single "transmission range" parameter, and the simulator does not natively capture contention on shared channels. Our choice of ONE therefore, prioritises scalability and DTN-specific modelling convenience over low-level physical fidelity.

As a result, the absolute values of performance metrics should be interpreted with caution, while relative comparisons between routing protocols under the same assumptions remain meaningful.

*C. Node groups and mobility model*

We model a total of 2 268 nodes divided into four host groups, representing different roles in the evacuation scenario: pedestrians, cars, emergency vehicles, and shelters. The chosen population of 2,268 simulated nodes is not intended to represent every resident of Hachijojima one-to-one.

The island has roughly 7,000 inhabitants [15], but only a subset of these people are expected to be present on the road network and able to act as DTN relays during the first 30 minutes after a tsunami warning. We therefore model an "active evacuation population" rather than the entire census population. The 1,500 pedestrians approximate a sizeable fraction of residents who evacuate on foot, including many elderly people who are less likely to rely on cars. The 750 vehicles represent households that use cars for evacuation, as well as traffic generated by people commuting or returning home when the earthquake occurs. The ten emergency vehicles capture a stylised presence of fire, rescue, and medical services circulating through the road network.

Finally, the eight shelters are selected from official hazard-mitigation maps for Hachijojima [16] and correspond to elevated evacuation sites. Using a reduced but still large population in this way keeps the simulation computationally tractable across multiple random seeds, while preserving the qualitative structure of contact opportunities between pedestrians, cars, emergency responders, and shelters. The number of nodes in each group is configured as

```
Scenario.nrofHostGroups = 4
Group1.nrofHosts = 1500
Group2.nrofHosts = 750
Group3.nrofHosts = 10
Group4.nrofHosts = 8
```

Group 1 corresponds to residents on foot. Their movement is constrained to the road network using ONE's MapBasedMovement model:

```
Group1.movementModel = MapBasedMovement
Group1.speed = 0.4,1.4
Group1.waitTime = 0,10
```

Walking speeds are drawn uniformly between 0.4 m/s and 1.4 m/s, covering frail elderly walkers up to relatively brisk walking. These values reflect the fact that Hachijojima had 7,042 residents in 2020, of whom 2,814 (40.0%) were aged 65 or older [15]. After reaching an intermediate destination on the map, nodes may pause for up to 10 seconds (e.g., due to congestion or hesitation) before moving again. The MapBasedMovement model selects random waypoints on the road graph and moves nodes along the connecting road segments, capturing uncoordinated, infrastructure-constrained movement without explicitly directing pedestrians towards shelters. Group 2 represents cars used for evacuation:

```
Group2.movementModel = MapBasedMovement
Group2.speed = 5,18
```

Vehicle speeds are drawn between 5 m/s and 18 m/s (approximately 18–65 km/h), reflecting a mixture of slow, congested traffic and faster movement on less crowded roads. Group 3 models fire and rescue vehicles:

```
Group3.movementModel = MapBasedMovement
Group3.speed = 10,20
```

with speeds between 10 m/s and 20 m/s (36–72 km/h), assuming that emergency vehicles can move faster and more purposefully than ordinary cars. Group 4 corresponds to shelters located on higher ground. These nodes are modelled as stationary:

```
Group4.movementModel = StationaryMovement
```

Specific shelter positions are configured using fixed node coordinates along safe road segments:

```
Group4.nodeLocation = 5622, 4593
Group4.nodeLocation = 5683, 4270
Group4.nodeLocation = 5565, 5385
Group4.nodeLocation = 4981, 5320
Group4.nodeLocation = 4824, 3848
Group4.nodeLocation = 4474, 3751
Group4.nodeLocation = 3871, 2854
Group4.nodeLocation = 4427, 2598
```

These locations were selected from the road network in areas identified as evacuation sites or higher-elevation zones in official hazard maps for Hachijojima [16]. As the ONE simulation cannot utilise decimals, rounding to the nearest integer will be employed on this occasion.

### D. Communication model and traffic generation

Wireless communication between nodes is modelled using two broadcast interfaces. All mobile and shelter nodes are equipped with a short-range, Bluetooth-like interface:

```
btInterface.type = SimpleBroadcastInterface
btInterface.transmitSpeed = 250k
btInterface.transmitRange = 30
```

The transmit speed of 250 kB/s corresponds to approximately 2 Mbit/s, and the 30 m range reflects a conservative estimate for device-to-device communication in urban and semi-urban environments.

Incidentally, we also experimented with a range below 20, but were unable to send a single message. Therefore, this time we shall assume the worst-case scenario and set it at 30 m range. Emergency vehicles in Group 3 additionally carry a higher-speed, medium-range interface to represent specialised radios or mobile mesh infrastructure:

```
highspeedInterface.type = SimpleBroadcastInterface
highspeedInterface.transmitSpeed = 10M
highspeedInterface.transmitRange = 50
Group3.nrofInterfaces = 2
Group3.interface1 = btInterface
Group3.interface2 = highspeedInterface
```

Pedestrians, cars, and shelters use only the Bluetooth-like interface.

```
GroupX.nrofInterfaces = 1
GroupX.interface1 = btInterface
```

Buffer capacities are chosen to reflect the differing storage resources of handheld devices, vehicles, and shelters:

```
Group1.bufferSize = 5M
Group2.bufferSize = 10M
Group3.bufferSize = 100M
Group4.bufferSize = 200M
```

Thus, shelters act as high-capacity information sinks, while evacuees operate under tighter storage constraints. Application-layer messages are generated using ONE's MessageEventGenerator:

```
Events.nrof = 1
Events1.class = MessageEventGenerator
Events1.interval = 40,60
Events1.size = 100k,500k
```

Messages are therefore created every 40–60 seconds, with sizes between 100 kB and 500 kB, representing short status updates, local observations, or small multimedia attachments. Sources are drawn uniformly from all mobile nodes in Groups 1–3, while all messages are addressed to shelters in Group 4:

```
Events1.hosts = 0,2260
Events1.tohosts = 2260,2268
Events1.prefix = S
```

With this configuration, node IDs 0–1499 correspond to pedestrians, 1500-2249 to cars, 2250-2259 to emergency vehicles, and 2260-2267 to the eight shelter nodes. Every message is assigned a time-to-live (TTL) of 30 minutes:

```
Group1.msgTtl = 30
Group2.msgTtl = 30
Group3.msgTtl = 30
```

which matches the simulated tsunami-warning period. Messages that are not delivered to a shelter before the end of the simulation are therefore considered expired. Shelters themselves do not generate messages in this scenario. Fig. 3 illustrates a snapshot of node positions during the evacuation, including pedestrians, vehicles and the eight shelter nodes.

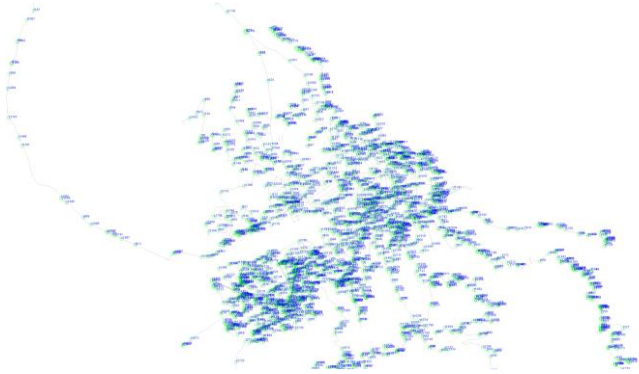

Fig. 3. The form of Hachijojima with node

### E. Routing protocols and configuration

We compare two DTN routing protocols implemented in ONE: EpidemicRouter and ProphetRouter. Both are evaluated under identical physical, mobility, and traffic conditions as described above; only the routing component is changed between experiments. For the Epidemic configuration, all groups use flooding-based routing:

```
Group1.router = EpidemicRouter
Group2.router = EpidemicRouter
Group3.router = EpidemicRouter
Group4.router = EpidemicRouter
```

For the PROPHET configuration, all groups instead use the probabilistic routing scheme:

```
Group1.router = ProphetRouter
Group2.router = ProphetRouter
Group3.router = ProphetRouter
Group4.router = ProphetRouter
```

We retain the default PROPHET parameters provided by ONE, including the time unit for ageing of delivery predictability (ProphetRouter.secondsInTimeUnit = 30) and the default values of α, β, and γ for initial predictability, transitivity, and ageing [10]. The focus of this work is on comparing the performance of flooding-based and history-aware routing under the same tsunami evacuation scenario, rather than on fine-grained parameter tuning of PROPHET.

### F. Experimental procedure

For each routing protocol, we run the simulator with the configuration described above. The EpidemicRouter and ProphetRouter are tested separately using otherwise identical settings. To mitigate the impact of randomness in node placement, message generation times, and contact patterns, each scenario configuration is executed multiple times with different random seeds. For each protocol, we aggregate the results over several runs and report average values (and, where appropriate, standard deviations) for all metrics. This experimental design allows us to isolate the effect of the routing protocol while keeping the underlying evacuation dynamics, mobility patterns, and traffic load constant. The resulting comparison highlights the trade-offs between reliability (delivery probability and latency) and efficiency (overhead, buffer usage, and message drops) for Epidemic and PROPHET routing in a realistic tsunami evacuation scenario on Hachijojima.

In practice, we executed each protocol configuration with five different random seeds (MovementModel.rngSeed ∈ 1, …,5). These seeds affect the initial placement of nodes, the realisation of walking speeds and waiting times, and the exact timing of message generation events. For each seed and each protocol we collected the full set of delivery and delay statistics from ONE's reports. We then computed the mean and, where appropriate, the standard deviation of each metric (delivery probability, end-to-end delay, overhead ratio, and average hop count) across the five runs. Using multiple seeds in this way reduces the risk that our conclusions are driven by a single particularly favourable or unfavourable random realisation of the evacuation dynamics. While the simulation model captures several key aspects of tsunami evacuation on Hachijojima, it also makes simplifying assumptions that limit the generality of the results. First, evacuee mobility is represented by independent shortest-path movements on a static road network with simple pause times. We do not model congestion, road closures due to damage, or changes in behaviour as people receive updated information, all of which could significantly affect contact opportunities in reality. Second, all devices are assumed to be willing participants in the DTN, with sufficient battery power and storage to remain active for the full 30-minute period. In practice, some fraction of users may have devices switched off, low battery, or privacy concerns that affect participation. Third, shelters are assumed to be stable, well-provisioned nodes with large buffers and no risk of failure, whereas real shelters might themselves suffer from power outages or connectivity problems. These limitations mean that our results should be interpreted as indicative of relative trends between routing protocols under a plausible, but idealised, scenario, rather than as exact forecasts of absolute performance in a real tsunami event.

## IV. SIMULATION RESULTS

### A. Performance evaluations

The effectiveness of each routing protocol is assessed using standard DTN performance metrics computed by ONE's reporting framework. The primary measures are delivery probability, average end-to-end delay, overhead ratio, average hop count, and the number of dropped messages [11] [14]. Delivery probability is defined as the fraction of generated messages that are successfully delivered to their intended destination before their TTL expires [17].

$$P_{del} = \frac{N_{delivered}}{N_{created}} \quad (1)$$

where, $P_{del}$ is Delivery probability, $N_{delivered}$ is number of delivered messages, $N_{created}$ is number of created messages.

The mean time between message creation and successful delivery, computed over all delivered messages. The ratio of the number of relayed (i.e., forwarded) message copies to the number of successfully delivered messages, minus one. This reflects the extra transmissions required to achieve successful delivery [17]:

$$Overhead = \frac{N_{relayed} - N_{delivered}}{N_{delivered}} \quad (2)$$

where, $N_{relayed}$ is relayed messages, $N_{delivered}$ is delivered messages. The average number of hops a message traverses between source and destination, indicating how widely messages are replicated through the network. The average hop count measures how widely messages are replicated through the network, and is computed as

$$H_{avg} = \frac{1}{N_{delivered}} \sum_{i=1}^{N_{delivered}} h_i \quad (3)$$

where $h_i$ is the number of hops (relay operations) traversed by the i-th delivered message. Finally, we consider the total number of dropped messages, i.e., messages that are removed from buffers due to TTL expiry or buffer overflow. This metric provides additional insight into the resource efficiency and failure modes of each protocol. All of these metrics are obtained from ONE's MessageStatsReport and per-message delivery logs generated at the end of each simulation run.

## B. Performance comparison

Table 1 compares the performance of Epidemic and PROPHET routing in the tsunami evacuation scenario with a 30m Bluetooth-like transmission range. Epidemic delivers roughly 40% of all generated messages to shelters within the 30-minute simulation window ($P_{deliv}$ = 0.4054), whereas PROPHET manages to deliver only about 3% ($P_{deliv}$ = 0.027). In absolute terms, this corresponds to roughly fifteen successfully delivered messages for Epidemic versus around one for PROPHET (out of 37 created messages). The average delivery delay of successful Epidemic messages is approximately 982s, about 16 minutes after the message creation, while the single message delivered by PROPHET arrives after roughly 555 s. This indicates that, when PROPHET happens to find a path, it tends to use relatively short routes. Consistently, the average hop count for Epidemic is much higher (14.73 hops) than for PROPHET (2 hops), reflecting the flooding-based nature of Epidemic routing. However, this comes at the cost of extreme replication: the overhead ratio for Epidemic is above 3000, meaning that on average more than three thousand relay operations are performed per successfully delivered message. Surprisingly, PROPHET does not reduce overhead in this scenario; its overhead ratio is even higher (3566), because a similar number of relay results in only a single delivery. Fig. 4 to Fig. 7 shows the results which have been plotted on separate graphs.

TABLE I. SIMULATION RESULTS FOR DIFFERENT RANDOM SEEDS

| Protocol | Seed | Delivery_prob | Avg. delay [s] | Overhead_ratio | Avg. hops | Dropped messages |
|---|---|---|---|---|---|---|
| Epidemic | 1 | 0.4054 | 981.8467 | 3020.3333 | 14.7333 | 11649 |
|  | 2 | 0.3784 | 937.5214 | 3044.8571 | 9.6429 | 11094 |
|  | 3 | 0.2973 | 834.7455 | 6219.1818 | 12.5455 | 33876 |
|  | 4 | 0.2973 | 964.9 | 4942.5455 | 11.6364 | 23523 |
|  | 5 | 0.2703 | 794.18 | 5760.2 | 11.1 | 26303 |
| PROPHET | 1 | 0.0270 | 555.4 | 3566 | 2 | 0 |
|  | 2 | 0.1351 | 848.12 | 629.2 | 5.2 | 0 |
|  | 3 | 0.1081 | 456.175 | 1373 | 3.5 | 0 |
|  | 4 | 0.027 | 324.3 | 4558 | 2 | 0 |
|  | 5 | 0.1351 | 892.46 | 966.2 | 3.4 | 0 |

TABLE II. SUMMARY OF DELIVERY PROBABILITY AND DELAY OVER DIFFERENT RANDOM SEEDS

| Protocol | Delivery_prob_mean | Delivery_prob_sd | Avg_delay_mean [s] | Avg_delay_sd [s] |
|---|---|---|---|---|
| Epidemic | 0.32974 | 0.052402389 | 902.63872 | 74.48483834 |
| PROPHET | 0.08646 | 0.049539826 | 615.291 | 221.1843544 |

TABLE III. SUMMARY OF OVERHEAD RATIO AND HOP COUNT OVER DIFFERENT RANDOM SEEDS

| Protocol | Overhead_ratio_mean | Overhead_ratio_sd | Avg_hops_mean | Avg_hops_sd |
|---|---|---|---|---|
| Epidemic | 4597.42354 | 1341.561351 | 11.93162 | 1.687794146 |
| PROPHET | 2218.48 | 1555.507411 | 3.22 | 1.183891887 |

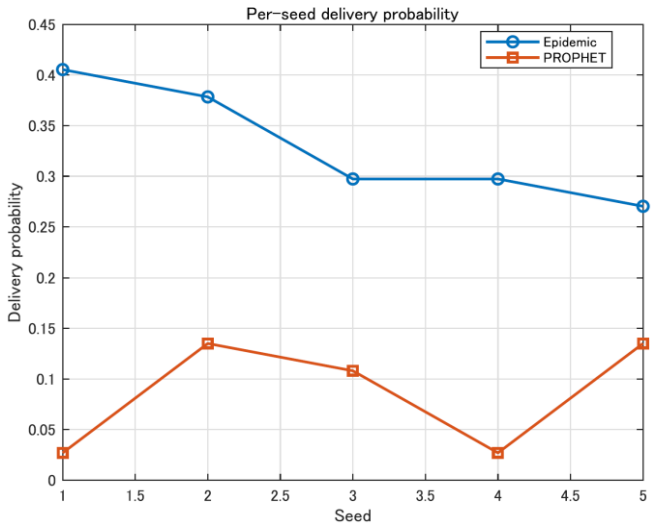

Fig. 4. Delivery probability of Epidemic and PROPHET

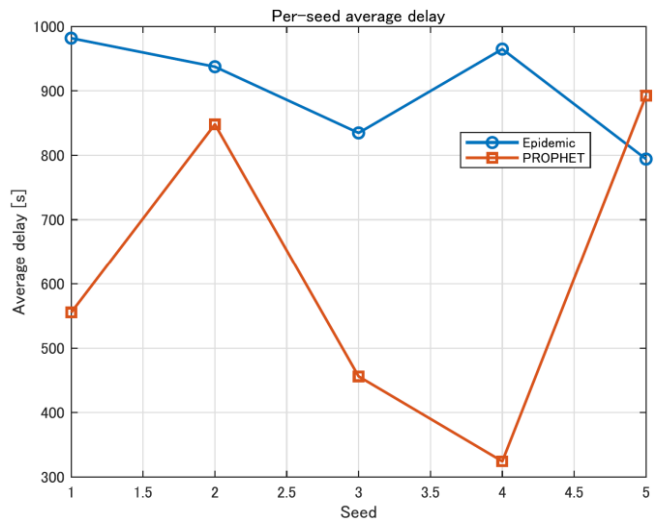

Fig. 5. Average delay under Epidemic and PROPHET

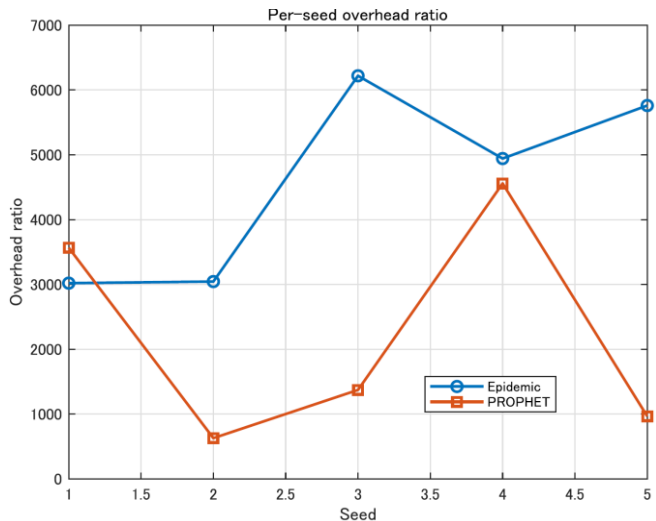

Fig. 6. Overhead ratio of Epidemic and PROPHET

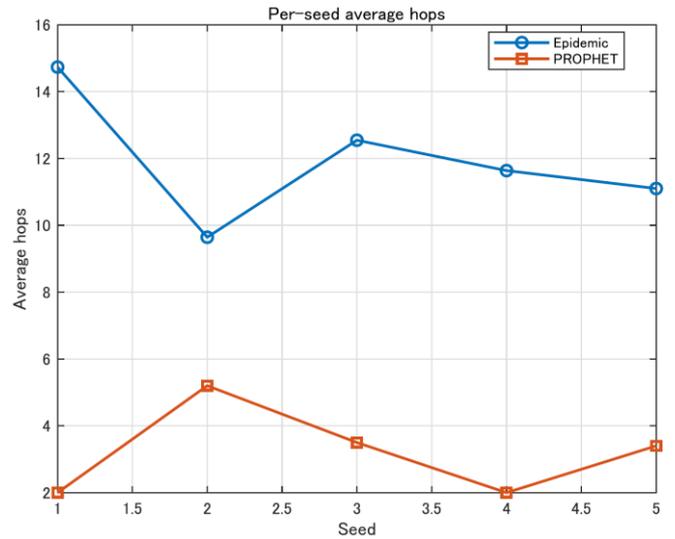

Fig. 7. Average hop count for Epidemic and PROPHET

## C. Discussion

Table 1 and the summary statistics in Table 2 and 3 report the performance of Epidemic and PROPHET routing over five independent runs with different random seeds. Overall, Epidemic achieves a substantially higher delivery probability than PROPHET, at the cost of larger overhead and longer paths. Across five seeds, Epidemic attains an average delivery probability of 0.33 with a standard deviation of approximately 0.05. Thus, roughly one third of all messages generated by evacuees reach one of the eight shelters within the 30-minute window, despite the sparse, fragmented contact graph.

PROPHET, in contrast, achieves only 0.086 mean delivery probability with similar variation (sd $\approx$ 0.05), so fewer than 10% of messages reach a shelter on average. This confirms that, although PROPHET does succeed in delivering some messages in most runs, its reliability under short, highly dynamic tsunami conditions is limited compared to flooding. In terms of latency, PROPHET performs better on average. Epidemic's mean delay for successfully delivered messages is about 903 s ($\approx$ 15 minutes) with a standard deviation of 74 s, meaning that most deliveries occur in the second half of the 30-minute evacuation window.

PROPHET's average delay is approximately 615 s ($\approx$ 10 minutes), but with a much larger standard deviation ($\approx$ 221 s). This indicates that, when PROPHET eventually finds a path to a shelter, it often does so via relatively short routes, but its performance is highly sensitive to the particular realisation of node contacts. The overhead statistics highlight the fundamental trade-off between robustness and resource consumption. Epidemic exhibits a mean overhead ratio of about 4,600, with considerable variation across seeds. This confirms that flooding generates a very large number of redundant copies per delivered message. PROPHET reduces the average overhead to roughly 2,200, i.e., about half of Epidemic's overhead on average, but still generates a significant number of relays given its much lower delivery probability. The average hop count mirrors this pattern: Epidemic paths are long (mean of 11.9 hops), while PROPHET paths are short (mean of 3.2 hops) but less frequently successful. It is also important to interpret these results in light of the earlier experiments with shorter transmission ranges. When the Bluetooth-like range was reduced to 15 m, no messages were delivered to any shelter within 30 minutes under either protocol, and latency-related metrics were undefined. This demonstrates that, under our map and mobility assumptions, there exists a critical range threshold below which shelters become effectively isolated components of the contact graph. Our main experiments therefore focus on the more realistic 30 m range, where a non-zero fraction of evacuees can reach shelters opportunistically. Taken together, the results show that Epidemic offers significantly higher reliability for life-critical tsunami evacuation at the cost of heavy resource usage, while PROPHET provides lower delay and reduced overhead but fails to deliver the majority of messages to shelters within the limited time frame.

Beyond these aggregate metrics, dropped messages reveal complementary failure modes of the two protocols. In ONE's MessageStatsReport, Epidemic drops between 11 094 and 33 876 messages per run due to buffer overflow, with a mean of 21 289, whereas PROPHET does not drop a single message in any of the five runs. This confirms that Epidemic's aggressive replication quickly fills buffers on mobile nodes and forces older messages to be discarded. PROPHET's conservative replication, on the other hand, keeps buffer pressure low enough that no buffer overflows occur. Given that both the message TTL and the simulation duration are 1 800 seconds, messages that are not delivered under PROPHET simply fail to encounter a viable path to a shelter within the 30-minute evacuation window; in a real deployment, such messages would soon be removed when their TTL expires. This contrast reinforces that it is not enough to look at delivery probability in isolation routing schemes that appear efficient in terms of overhead and buffer usage may still fail to deliver a large fraction of critical messages before they expire.

## V. CONCLUSION AND FUTURE WORKS

### A. Conclusion

In this paper, we studied the use of delay-tolerant networking to support tsunami evacuation on Hachijojima, using a map-based scenario in the ONE simulator with pedestrians, cars, emergency vehicles, and stationary shelters. We focused on a 30-minute warning window during which existing infrastructure is assumed to be unavailable, and all communication relies on opportunistic contacts over a 30 m Bluetooth-like interface. Messages are generated by evacuees in the field and addressed to shelters that act as high-capacity information sinks. Our multi-run evaluation compared Epidemic and PROPHET routing under identical physical and traffic conditions. On average over five seeds, Epidemic delivered roughly one third of all messages to shelters ($P_{deliv} \approx 0.33$), with mean latency of about 15 minutes and very high overhead and hop counts. PROPHET, by contrast, delivered less than 10% of messages on average ($P_{deliv} \approx 0.086$), albeit with lower mean delay (about 10 minutes), roughly half the overhead, and much shorter paths. Experiments with a reduced transmission range below 20 m transmission range further showed that physical connectivity constraints can completely prevent shelters from receiving any messages within the evacuation window. These findings suggest that, for the initial 30-60 minutes of tsunami evacuation on a small island, history-based probabilistic routing faces inherent limitations. PROPHET depends on contact histories that have little time to stabilise in a rapidly evolving emergency, and its conservative forwarding policy can miss many potential paths towards shelters. Flooding-based routing such as Epidemic is much more robust in this regime: by replicating messages aggressively, it significantly increases the chance that at least one copy eventually reaches a shelter, even in a sparse and partitioned network. From a public safety perspective, delivery probability is arguably the dominant objective, and our results indicate that a controlled form of flooding is better aligned with this goal than purely probabilistic forwarding. At the same time, the very high overhead of Epidemic raises practical concerns. In a real deployment, finite buffers, energy constraints, and contention at the wireless channel would limit the number of replicas that can be sustained. Our results therefore support a design strategy in which a robust flooding-like mechanism is used during the critical early phase of evacuation, but is combined with simple control mechanisms such as copy limits, prioritisation of messages destined for shelters, or preferential forwarding by emergency vehicles to keep resource usage within acceptable bounds.

### B. FutureWorks

Future work will extend this study in several directions. First, the current mobility model treats evacuees as uninformed agents who move along the road network without explicit knowledge of shelter locations. Incorporating behavioural models where evacuees gradually learn about shelters and adjust their routes accordingly would enable a more realistic assessment of how information and movement co-evolve. Second, we plan to evaluate additional routing protocols, including quota-based schemes such as Spray-and-Wait and buffer-aware strategies such as MaxProp, CafRepCache [18] and CognitiveCache [19], to explore whether they can approximate the robustness of Epidemic while reducing overhead. Third, a more systematic exploration of PROPHET parameter settings, including ageing time and transitivity coefficients, could clarify how much of its poor performance is due to fundamental limitations versus suboptimal tuning for short-term evacuation scenarios. Finally, we intend to incorporate energy aware opportunistic charging and energy distribution as in SmartCharge [20], heterogeneous radio technologies, and more detailed shelter and population data from real-world tsunami preparedness plans for Hachijojima and similar islands. Such extensions would move the model closer to operational deployment and help identify practical configurations for DTN-based tsunami evacuation support systems.